\documentclass[letterpaper, 10 pt, conference]{ieeeconf}  

\usepackage{booktabs}
\usepackage{nicematrix}
\usepackage{graphicx}

\IEEEoverridecommandlockouts                              

\title{\LARGE \bf
Agent-based Simulation Evaluation of CBD Tolling: A Case Study from New York City
}

\author{Qingnan Liang$^{1,2}$, Ruili Yao$^{3}$, Ruixuan Zhang$^{1}$, Zhibin Chen$^{2,4}$, Guoyuan Wu$^{5}$
\thanks{*Financial support from the National Science Foundation under Grant DGE-2152258 (M. Barth, PI) is gratefully acknowledged.}
\thanks{$^{1}$Qingnan Liang and Ruixuan Zhang are with the Department of Civil and Urban Engineering, 
New York University, New York, United States.
        }%
\thanks{$^{2}$Qingnan Liang and Zhibin Chen are with Shanghai Key Laboratory of Urban Design and Urban Science,
        NYU Shanghai, Shanghai, China.
        }%
\thanks{$^{3}$ Ruili Yao (Corresponding author) is with the Department of Electrical and Computer Engineering, 
University of California, Riverside, Riverside, California, United States.
        {\tt\small ruili.yao@email.ucr.edu}}%
\thanks{$^{4}$ Zhibin Chen is with Shanghai Frontiers Science Center of Artificial Intelligence and Deep Learning,
        NYU Shanghai, Shanghai, China.
        }%
\thanks{$^{5}$Guoyuan Wu is with the Center of Environmental Research and Technology,
University of California, Riverside, Riverside, California, 
United States.}%
}

\begin{document}

\maketitle
\thispagestyle{empty}
\pagestyle{empty}

\begin{abstract}

Congestion tollings have been widely developed and adopted as an effective tool to mitigate urban traffic congestion and enhance transportation system sustainability. Nevertheless, these tolling schemes are often tailored on a city-by-city or even area-by-area basis, and the cost of conducting field experiments often makes the design and evaluation process challenging. In this work, we leverage MATSim, a simulation platform that provides microscopic behaviors at the agent level, to evaluate performance on tolling schemes. Specifically, we conduct a case study of the Manhattan Central Business District (CBD) in New York City (NYC) using a fine-granularity traffic network model in the large-scale agent behavior setting. The flexibility of MATSim enables the implementation of a customized tolling policy proposed yet not deployed by the NYC agency while providing detailed interpretations. The quantitative and qualitative results indicate that the tested tolling program can regulate the personal vehicle volume in the CBD area and encourage the usage of public transportation, which proves to be a practical move towards sustainable transportation systems. More importantly, our work demonstrates that agent-based simulation helps better understand the travel pattern change subject to tollings in dense and complex urban environments, and it has the potential to facilitate efficient decision-making for the devotion to sustainable traffic management.

\end{abstract}


\section{Introduction}

Congestion profoundly affects living quality and future sustainability, primarily through travel delays, increased energy consumption, and elevated air pollution. Travel delays reduce individual productivity and life satisfaction due to longer commutes and lead to broader economic inefficiencies. Higher energy consumption, particularly of fossil fuels, occurs as vehicles idle in traffic, which can have significant environmental and geopolitical impacts. The increased air pollution from congested traffic contributes to serious health issues and diminishes public health. In terms of sustainability, congestion ruins the efforts towards greener, more sustainable cities by promoting reliance on non-renewable energy and hindering the adoption of eco-friendly transportation solutions.

Tolling can be pivotal in managing and alleviating traffic congestion in urban areas. Through dynamic pricing, which adjusts toll rates based on real-time traffic conditions, tolling can incentivize drivers to travel during less congested, off-peak hours or search for alternative routes. This approach helps evenly distribute traffic and contributes to a reduction in overall vehicle emissions. However, designing an optimal tolling policy that achieves the desired outcomes remains a significant challenge, particularly in large cities. Field experiments, often necessary for such endeavors, can be prohibitively expensive and complex. Therefore, simulation emerges as a low-cost and flexible alternative. Traditional simulation models, however, fall short as they often oversimplify real-world complexities. More advanced simulation tools are needed to accurately mirror the complex dynamics of urban traffic and aid in formulating effective toll road policies.

To address the abovementioned issues, we need to figure out a simulation method to provide details and realism to the real world.
With various techniques, agent-based simulations can be implemented as a high-resolution enabled modeling framework. Due to the individual tracking characteristics, agent-based simulations have a more accurate, explainable, and reliable real-world modeling effect. Sometimes, such simulations were realized as discrete event-based, Monte Carlo, or microscopic transport simulations. Due to the complexity of users' behavior and traffic dynamics, researchers typically adapt existing simulation frameworks for transport simulations. Among those simulation frameworks, the MATSim modeling framework is prevalent for its mature and complete framework development. MATSim is an open-sourced agent-based transport simulation framework implemented in JAVA. It has been validated and used in many scientific and industrial research \cite{eth_zurich_multi-agent_2016}. 

On the other hand, the crucial traffic condition of NYC provides us with a perfect opportunity for a case study. New York is one of the top five congested urban areas in the United States. The Manhattan CBD in New York City is the core business and residential area. It is also the largest and most economically significant metropolitan region and commercial center in the United States. There are 1.5 million jobs, 450 million square feet of office space, and more than 617,000 residents. Due to the intense congestion conditions in NYC, the MTA plans to develop a tolling program to control the traffic volume in the Manhattan central business district (CBD).

In this work, we leverage MATSim, an agent-based simulation platform that provides fine granularity of microscopic agent behaviors, to evaluate performance on different tolling schemes. Specifically, we focus on the Manhattan Central Business District (CBD) in New York City (NYC). This research can help better understand the travel patterns in the dense urban environment and provide efficient decision-making support for sustainable transportation systems.

The rest of the paper is organized as follows. Section \ref{literature} conducts recent literature on congestion tolling and the simulation models. Section \ref{model} introduces an agent-based simulation framework, i.e., MATSim and the pricing extension model. Section \ref{scenarios} presents the scenarios for investigation, including the CBD tolling program for NYC and a build-up and well-calibrated NYC test bed. Section \ref{results} visualizes the travel demand when the Manhattan CBD is under the toll program and conducts the evaluation analysis. Section \ref{conclusion} concludes the paper and discusses the future work.   

\section{Literature Review} \label{literature}
\subsection{Traffic congestion tolling and modeling}
Congestion pricing policies significantly impact people's travel behaviors and can benefit multiple stakeholders of drivers, public transit, and governments \cite{small1992using}. Theoretically, the first-best (Pigouvian toll) pricing can shift travels toward a system optimal state \cite{yang1998principle}. Although the marginal cost pricing scheme has a perfect theoretical foundation, it could be more practical due to the issues ranging from operating cost and fairness to public acceptance.  The toll collection cost may justify not setting for the entire road network but only some of its primary congested links.  To this end, second-best road pricing schemes have been devised, e.g., the cordon-based road pricing scheme and the link-specific tolling scheme \cite{yang2010road}. 

Regarding the charging methods, there are five common ways for toll programs in the US according to the Federal Highway Administration of the US Department of Transportation, \cite{decorla2003value}: variably priced lanes, variable pricing on roadways, cordon charges, and area-wide pricing. For example,  the variably priced lanes pricing strategy is adopted by the I-15 interstate highway in San Diego \cite{brownstone2003drivers}, where dynamic tolls are charged for vehicles using the High Occupancy Vehicle (HOV) lanes, and the prices vary with different demand on the HOV lanes. Another example of adopting variable pricing on entire facilities can be found in Lee County in Florida \cite{burris1998planning}, where they examined two bridge structures with all lanes on them. Area-wide pricing has been tested by the State of Oregon, where a pricing scheme involving per-mile charges will be considered a replacement for fuel taxes in the future, with higher charges during congested periods on high-traffic road segments. Specifically, the NYC Department of Transportation (DOT) plans to propose a congestion pricing plan in the area of Manhattan in the southern part of 60th Street, and the expected revenue will be used to improve and maintain the public transit system in NYC \cite{baghestani2020evaluating}. What complicates things is that the congestion pricing policy should reflect the dynamic changes of time during a day, the traveler's elasticity, and the proportion of other modes.

\subsection{Agent-Based Models for Transport Simulation}

According to the officials in NYC, there are several policy tools or models to evaluate newly proposed policies and technologies \cite{he2021validated}. One is the New York Best Practice Model developed by the New York Metropolitan Transport Council, which is designed for long-term capital planning but not short-term quick response evaluation. Therefore, it may fail to reflect the dynamics of the evolution of transportation systems fully. Another tool is the Balanced Transportation Analyzer, developed by the Nurture Nature Foundation. However, this model fails to capture spatial-temporal dependencies within the city.

Recently, with the advancement of computational ability, agent-based models are receiving more and more attention. The agent in the system starts from the local rules to the more complex adaptive and emerging behaviors formed by the interaction of the neighborhood agents, thus yielding the system dynamics in the environment \cite{huang2022overview}. The concept of agent-based simulation is a good match with application in transportation management systems. There are several open-sourced software toolkit: MATSim \cite{w2016multi}, TRANSIMS \cite{smith1995transims}, GAMA \cite{drogoul2013gama}, SACSIM \cite{bradley2010sacsim}. 

Among these options, the MATSim modeling framework is prevalent for its mature and complete framework development. MATSim is an open-sourced agent-based transport simulation framework implemented in JAVA \cite{eth_zurich_multi-agent_2016}. It has been validated and used in many scientific and industrial research. For example, the MATSim model is used for assessing the impact of future electric vehicle charging infrastructure for long-distance transport in Sweden \cite{marquez2021assessment}, estimating public charging demand and optimizing charging station location at urban scale \cite{yi2023agent}, analyzing the spatial and temporal impact of COVID-19 on network performance for NYC in the post-pandemic era \cite{wang2021mobility} and so on. 

\section{MATSim Congestion Pricing Model} \label{model}

\begin{figure}[htbp]
    \centering
    \includegraphics[width=\linewidth]{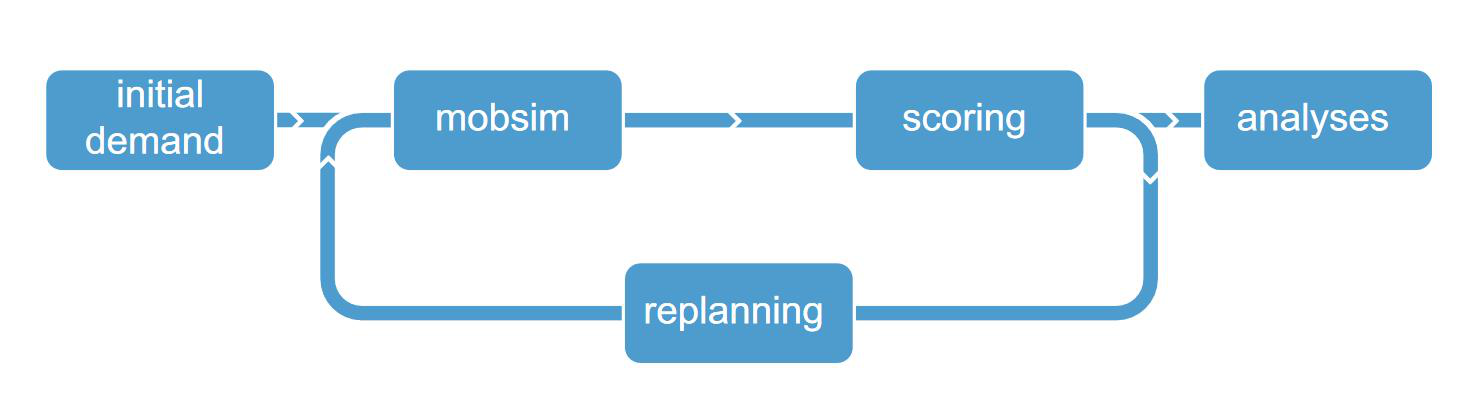}
    \caption{An illustration of MATSim loop}
    \label{matsim loop}
\end{figure}

In this section, we introduce the operating principles of MATSim to help us better understand the framework and know why it is suitable for serving as a test bed for scenario experiments.
Figure \ref{matsim loop} illustrates the MATSim simulation loop, a repetitive process integral to its internal mechanics. This loop simulates a synthetic population of agents in each cycle, focusing on their travel behavior. These agents generate their travel needs by forming plans that detail starting points, destinations, purposes, timings, route selections, and modes of transportation for each journey. These predefined travel plans are carried out in a simulated, realistic setting during the mobility simulation (mobsim). Through repeated cycles, MATSim enhances agent behavior using a co-evolutionary learning method. This method scores each individual's travel plan based on the effectiveness of the executed actions. Agents may then revise their plans based on these scores to improve outcomes. This cycle of scoring and replanning leads to heuristic optimization of the plans across several iterations, embodying an evolutionary strategy. This co-evolutionary method incorporates the interplay between agent behaviors and plan evaluations within their simulated context. The training observes the emergence of collective behaviors as agents interact within this virtual setting. This process continues until no agent can independently alter its behavior for better performance, a condition known in mathematical terms as a Nash equilibrium. This state is acknowledged for its accuracy in mirroring real-world traffic patterns. When a Nash equilibrium is reached in the simulation, it enables detailed reflection and analysis of system-wide and individual agent states.

What adds to MATSim's versatility is the extensive range of extensions and contributions from the MATSim community, broadening its applicability. These enhancements span various aspects like input data preparation, individual car traffic modeling, mode choice models, commercial traffic extensions, more choice dimensions, intra-day replanning, calibration techniques, visualization, and analysis tools. Researchers are also provided with guidelines for developing their unique extensions. They can tap into three levels of MATSim plugins: the main distribution, community contributions, and personal extensions. Of particular interest here is the ``road pricing" extension. This extension calculates and applies a toll for each vehicle entering a specific link at a certain time, with the charge assigned to the driver. The scoring formula is adjusted to incorporate this, as shown in the equation:
\begin{align*}
S_{trav,car,q}=\cdots+\beta_{m}\cdot \tau+\cdots,
\end{align*}
Here, $\tau$ represents the toll costs (usually a negative value) within the budget, and $\beta_{m}$ is the marginal utility of money. By altering the scoring function, drivers are influenced to consider toll costs in their decision-making process, affecting route choice, departure time, mode of transport, and destination. This extension allows for a more scientific and efficient evaluation of strategies like the Manhattan CBD tolling program.

\section{Description of The Scenarios} \label{scenarios}

\subsection{Future CBD Tolling Program in New York City}

\begin{figure}[t!]
\centering
\includegraphics[width=0.7\linewidth]{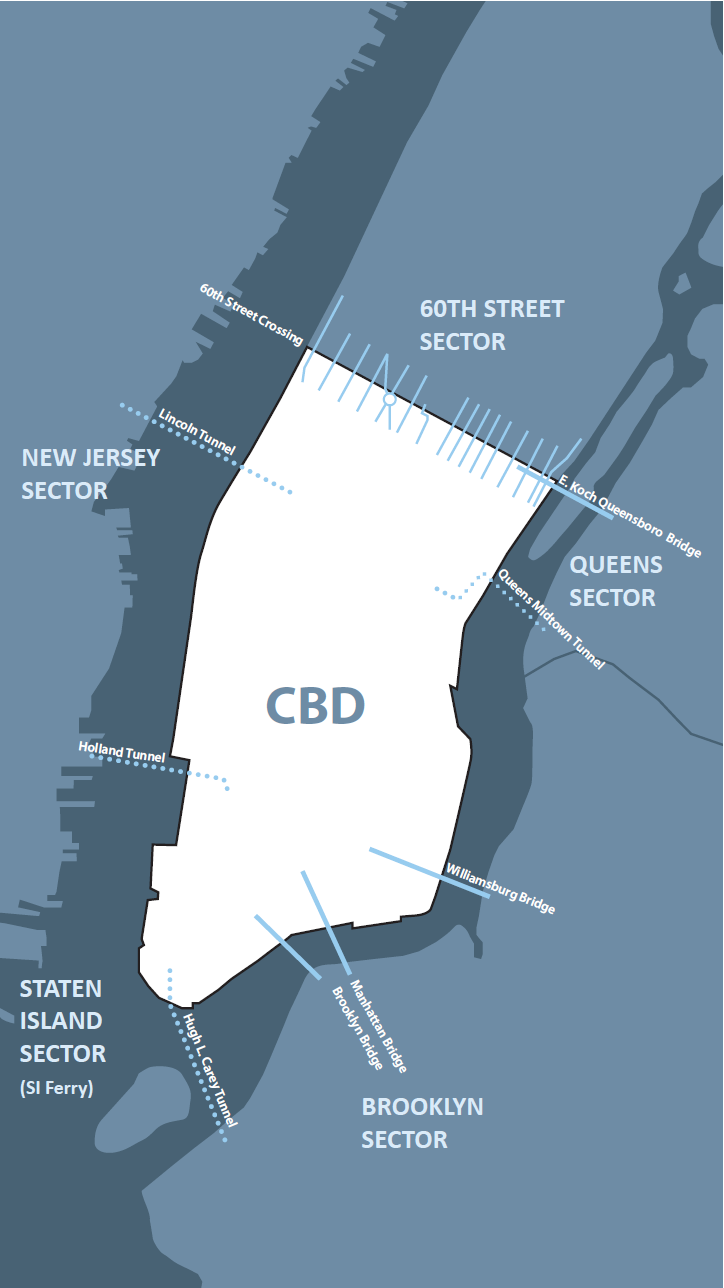}
\caption{Charging region of the tolling program}
\label{CBD maps}
\end{figure}

 In recent years, the New York state and city agents and stakeholders have been willing to solve the congestion issue in Manhattan CBD by charging tolls. In April 2019, the state enacted the MTA Reform and Traffic Mobility Act, which mandated the MTA's Bridge and Tunnel Authority to design, develop, construct, and operate a CBD tolling program. In addition, the federal government requires an environmental assessment of the potential environmental impact of the CBD tolling program. If the federal government approves the program, the environmental assessment process will include the implementation of solid public awareness. The CBD tolling program will charge passenger cars once daily for entering or leaving the CBD. The toll charge is changed by variable tolling, and no toll will be charged for qualifying authorized emergency vehicles and vehicles transporting people with disabilities. This program aims to reduce daily vehicle miles traveled within the Manhattan CBD by at least 5 percent, reduce the number of vehicles entering the Manhattan CBD daily by at least 10 percent, and create a funding source for the MTA Capital Program.  More details of tolling scenarios and prices will be discussed later in this article.
From the geographic view, the Manhattan CBD is composed of Manhattan's south part and includes 60th Street, excludes Franklin D. Roosevelt (FDR) Drive and the West Side Highway/ Route 9A, the Battery Park Underpass, and the surface roadway portion of the Hugh L. Carey Tunnel that connects to the West Side Highway/Route 9A. Figure \ref{CBD maps} shows the charging region of the tolling program. 

The authority proposed six alternatives for the tolling plan based on different purposes. We chose the base plan for the evaluation showcase, which is as follows: the authority charges \$ 9 for the peak hours (6 a.m. to 8 p.m.), \$ 7 for the off-peak hours (8 p.m. to 10 p.m.), and \$ 5 for overnight (10 p.m. to 6 a.m.). Next, we introduce the MATSim-NYC test bed to support our evaluation of the tolling programs. 

\subsection{MATSim-NYC Test Bed}
MATSim-NYC is an open-source multi-agent simulation model for New York City that is well calibrated and validated \cite{chow2020multi}. The road network is extracted from OpenStreetMap, and the transit schedule is obtained from GTFS. A representative synthesized population is generated from PopGen 2.0 \cite{konduri2016application}, where a population
of 8.24 million people is generated for the base year of 2016, compared to a total actual population of 8.34 million. There was an average of 4\% difference between the synthetic population and the Longitudinal Employer-Household Dynamics 2016 data. Travel agendas were replicated from the 2010/2011 Regional Household Travel Survey (RHTS), minus the mode for each trip. Over 30 million trips were synthesized. Modes were synthesized using a mode choice model. The mode choice model is a tutor-based logit model, and it is calibrated using 2010/2011 RTHS and named MATSim-NYC. The calibrated synthetic population and MATSim-NYC model can be the baseline compared to other customized scenarios like tolling scenarios. 

This test bed can model the heterogeneous population and complex transportation system, which can help us evaluate the impacts of city-scale policies on large cities like NYC.

\section{Evaluation Results} \label{results}

\begin{figure}[htbp] 
\centering
\includegraphics[width=\linewidth]{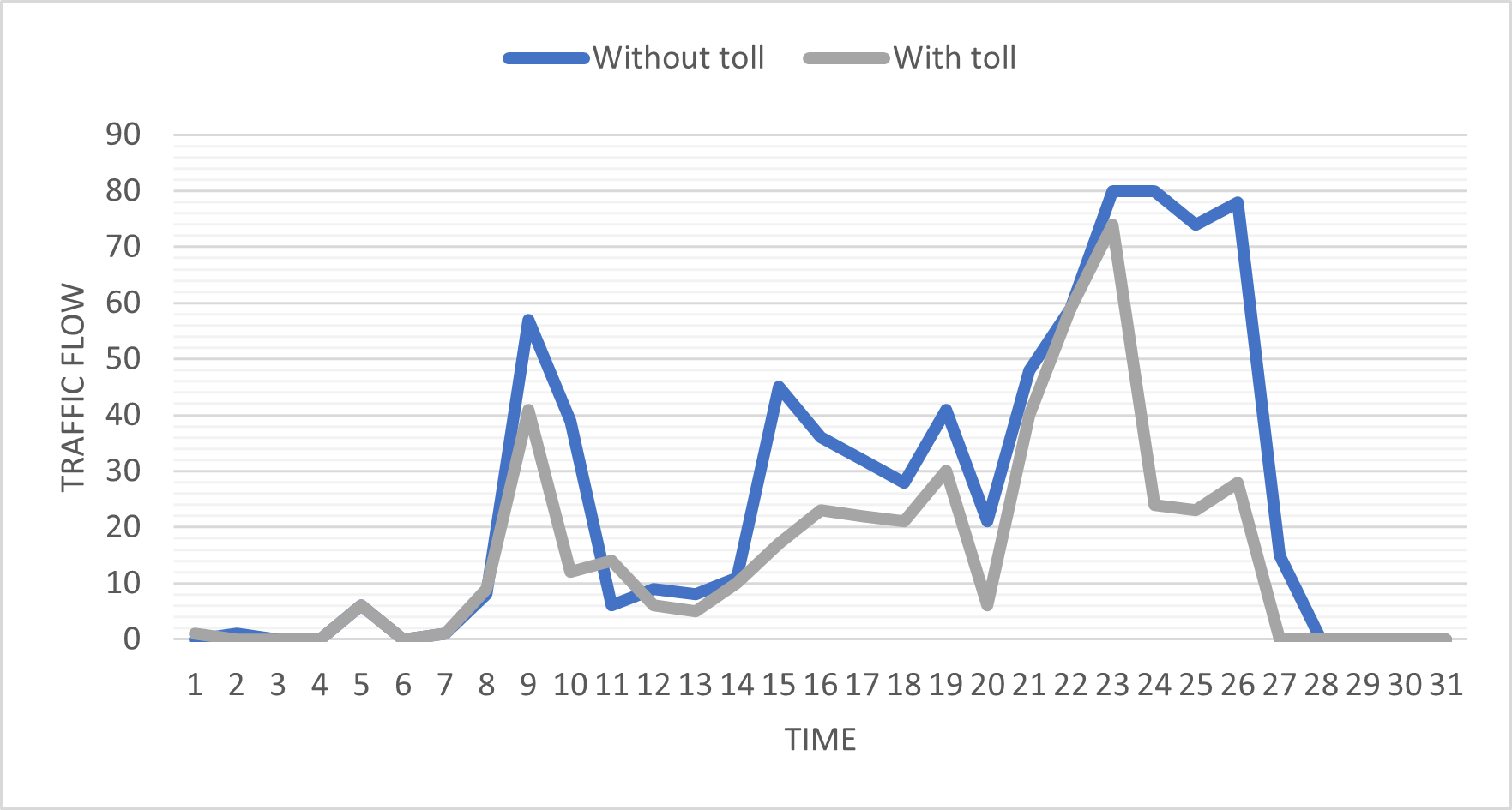}
\caption{The distribution of the link flow during the day}
\label{8a}
\end{figure}

\begin{figure}[htbp]
\centering
\includegraphics[width=\linewidth]{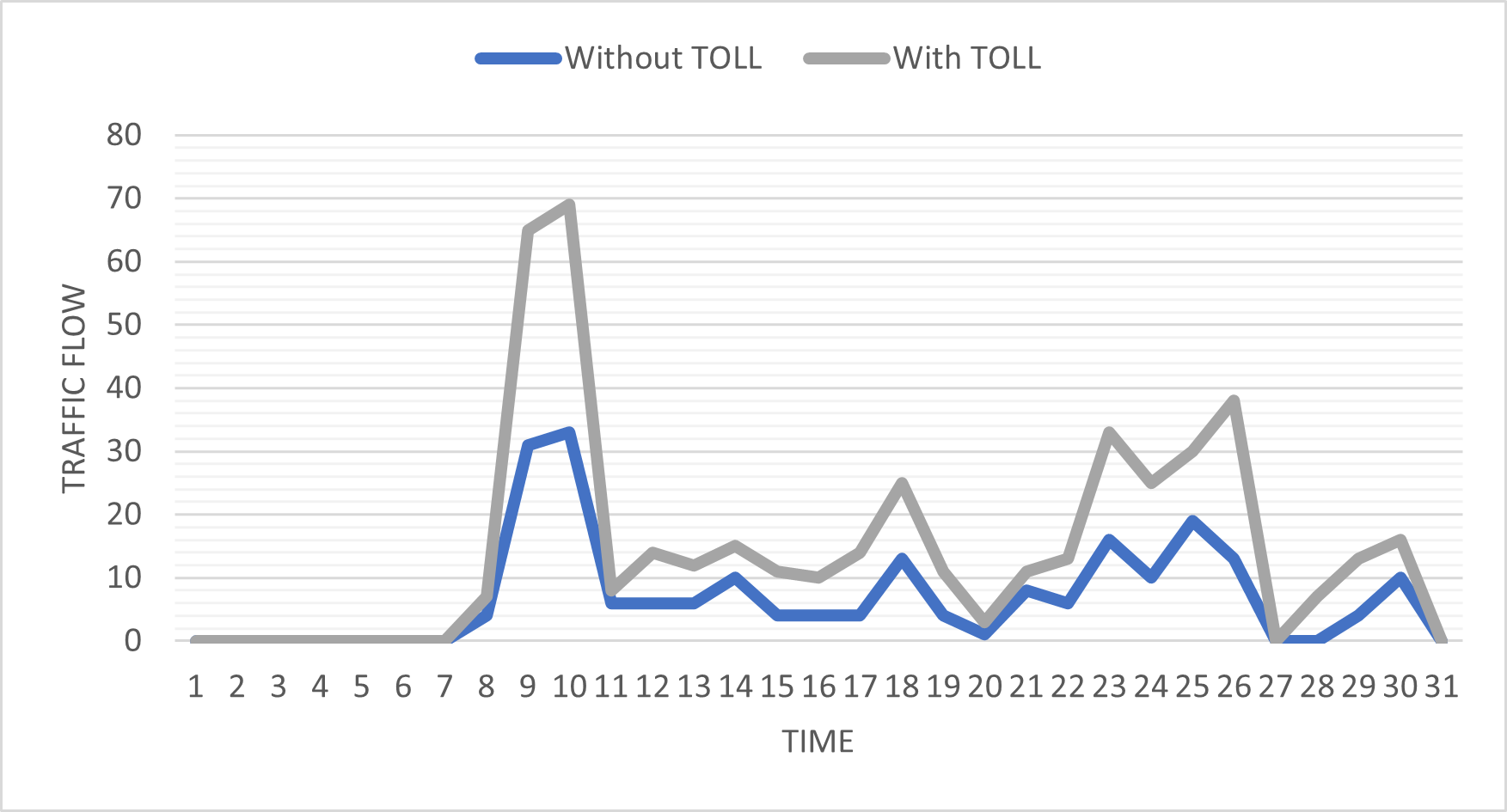}
\caption{The distribution of the link flow during the day}
\label{8b}
\end{figure}

We use the pricing extension from the MATSim contribs to set up the pricing plan for the simulation and then evaluate the tolling effect of the program strategy.
We first compare the traffic volume of the traffic links under the cases before and after the toll program was applied in Manhattan CBD. Figure \ref{8a} - \ref{8b} shows two typical distributions of the link flow. The grey line represents the traffic volume after the toll charge, and the blue line represents the original traffic volume in the Manhatten CBD area.  As we can see, the effect of the tolling program on the specific link is not determined. For example, we can observe that the traffic volume on some links is depressed by the pricing scheme, as shown in Figure \ref{8a}, but some volume increases, as shown in Figure \ref{8b}. Besides, the volume on some links doesn't change too much. 

\begin{figure}[htbp]
\centering
\includegraphics[width=\linewidth]{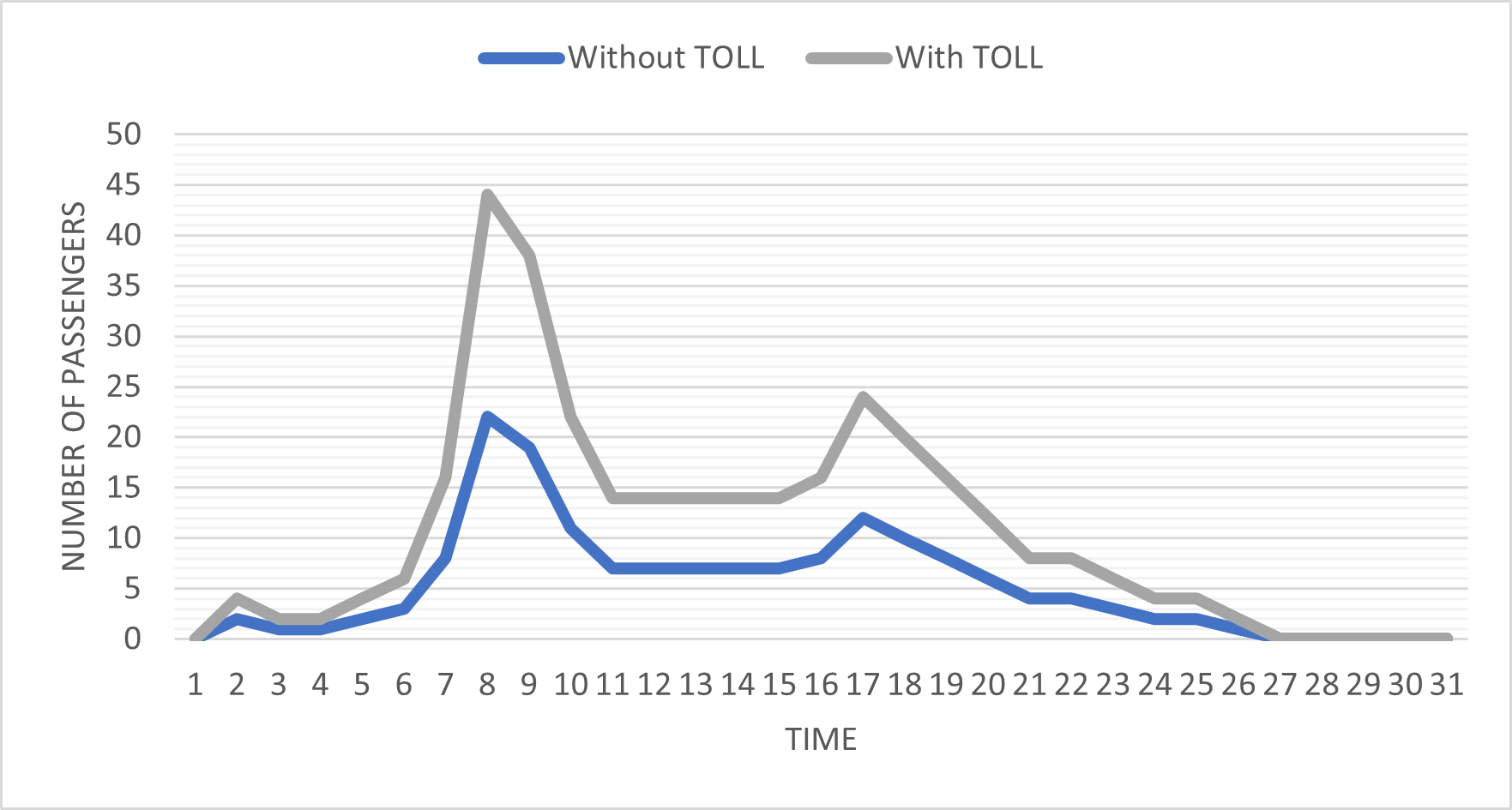}
\caption{The distribution of the public transit passengers during the day}
\label{11a}
\end{figure}

\begin{figure}[htbp]
\centering
\includegraphics[width=\linewidth]{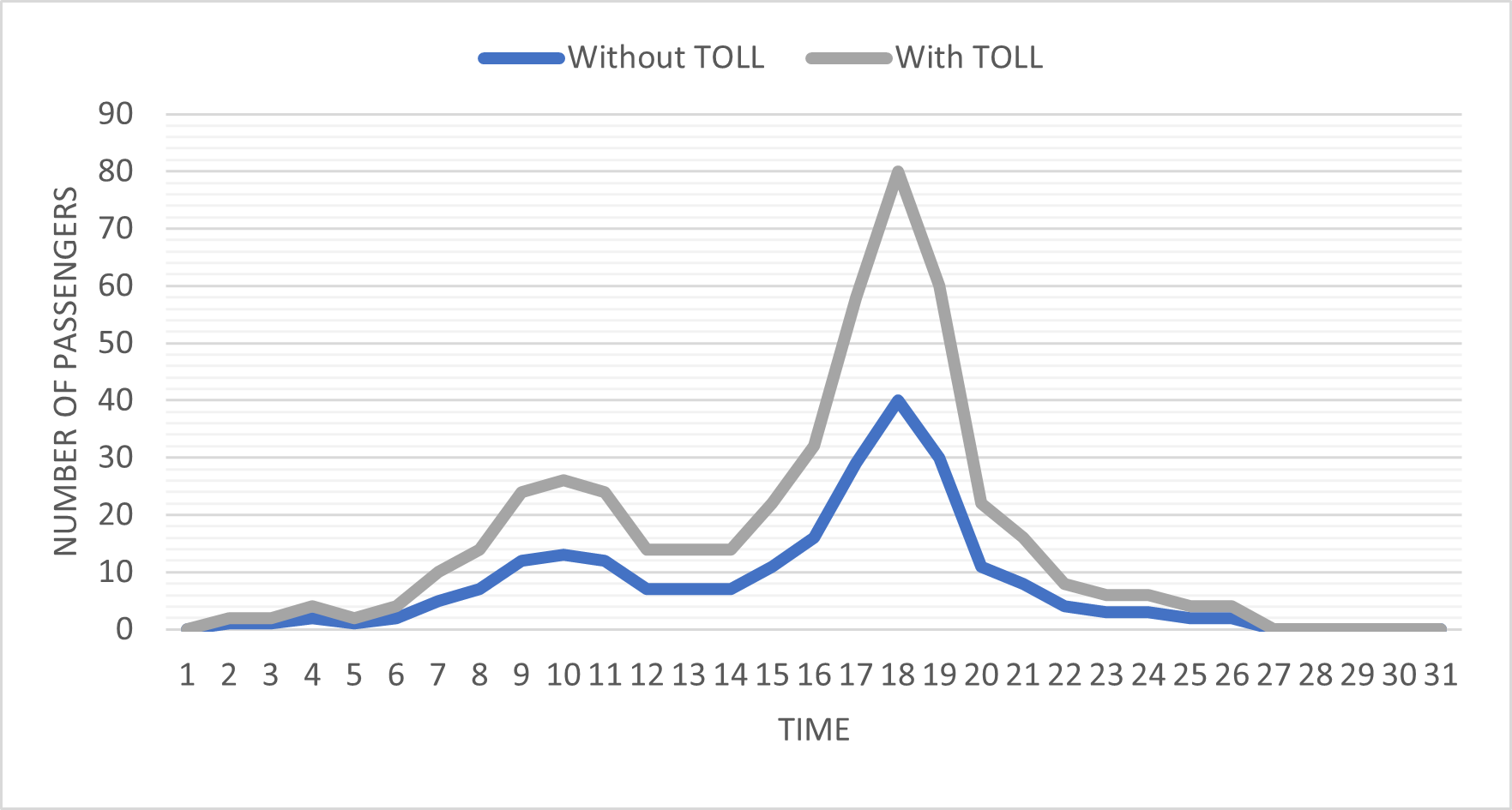}
\caption{The distribution of the public transit passengers during the day}
\label{11b}
\end{figure}

Meanwhile, we compare the traffic data on public transportation after the toll program was applied in Manhattan CBD. The grey line represents the traffic volume after the toll charge, and the blue line represents the original traffic volume in the Manhatten CBD area. The result is shown in 
Figure \ref{11a} - \ref{11b}. The observation is rather obvious. After implementing the tolling scheme, some agents switched to public transit to avoid the congestion price on private cars. This is reasonable and easy to deduce. Notably, there are many more switches during commuting hours. So we can know that the pricing program will reshape the citizen's travel modes to work.

Furthermore, based on output data, we can compare the mode choice and trip score before and after adding the pricing strategy. Firstly, the mode choice is shown in Table \ref{table: mode choice}, where we notice that after applying the designed pricing strategy, few agents use cars while more prefer non-car modes, such as bike and public transit. While the ratio is not significant (and we don't want to see a dramatic change as well), this influence of the pricing strategy is still demonstrated to be positive in balancing the mode choice to reduce traffic congestion. 

Besides balancing the mode choice, the trip score in such an agent-based simulation indicates the success rate that one agent can complete its route: the higher the score value, the more satisfied the agent feels about its trip. From our simulation results, we compare the distribution of the trip scores of all agents in two scenarios, as shown in Table \ref{table: trip score}. It can be noticed that with pricing, the mean of trip scores increase by 0.28 (7.8\%) while the standard deviation and the maximum value decrease, and the minimum value increases a little bit. It can be interpreted that the proposed pricing strategy improves the trip score distribution by reducing the standard deviation and raising the minimum value while maintaining a similar distribution for those already positive scores.

\begin{table}[htbp]
\footnotesize
\centering
 \caption{Change in mode choice}
\label{table: mode choice}
  \begin{tabular}{c|c c c c}
	\hline
\bf{Scenario} & FHV & Access walk & Bike & Car  \\
	\hline
Without pricing & 24323 & 436050 & 20361 & 369908  \\
 With pricing & 25678 & 438878 & 20517 & 363878  \\
 Change ratio \%& 5.57 & 0.65 & 0.77 & -1.63  \\
	\hline
\bf{Scenario}   & Pt & Ride & Taxi & Transit walk \\
	\hline
Without pricing & 956534 & 42741 & 39095 & 424597  \\
 With pricing & 963930 & 43250 & 39283 & 428541 \\
  Ratio & 0.77 & 1.19 & 0.48 & 0.93 \\
  
	\hline
 \bf{Scenario}   & Cb & Egress walk &Walk\\
	\hline
Without pricing & 1404 & 436050  & 288564 \\
 With pricing & 1319 & 438878 & 289344 \\
  Ratio & -6.05 & 0.65 & 0.27 \\
	\hline
 \end{tabular}
\end{table}

\begin{table}[htbp]
\caption{Change in trip score}
\centering
\NiceMatrixOptions{notes/para,notes/enumitem-keys-para={itemjoin = ;\;}}
\begin{NiceTabular}{c c c c c c }
    \toprule
    \RowStyle{\bfseries}
    Scenario & Mean & Std  & Minimum & Maximum  & Median\\ \midrule

    Without pricing & 3.60 & 48.32 & -567.95 & 61.55  & 19.48 \\
    With pricing & 3.88 & 47.75 & -565.99 & 61.52  & 19.47 \\
    Difference & 0.28 & -0.57 & 1.96 & -0.03  & -0.01 \\
\bottomrule
\end{NiceTabular}
\label{table: trip score}
\end{table}

\begin{figure}[htbp]
\centering
\includegraphics[width=\linewidth]{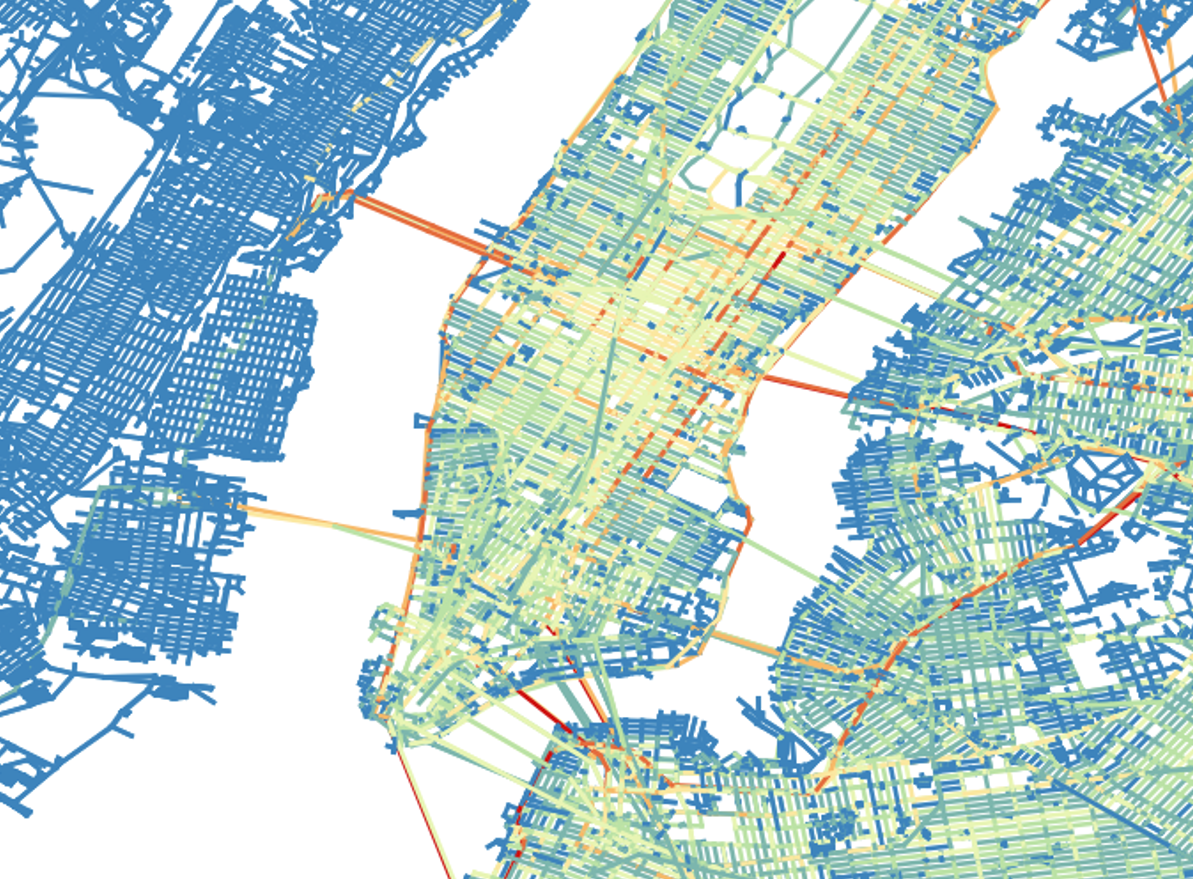}
\caption{Link congestion level during 17:00 - 18:00 without toll}
\label{withouttoll}
\end{figure}

\begin{figure}[htbp]
\centering
\includegraphics[width=\linewidth]{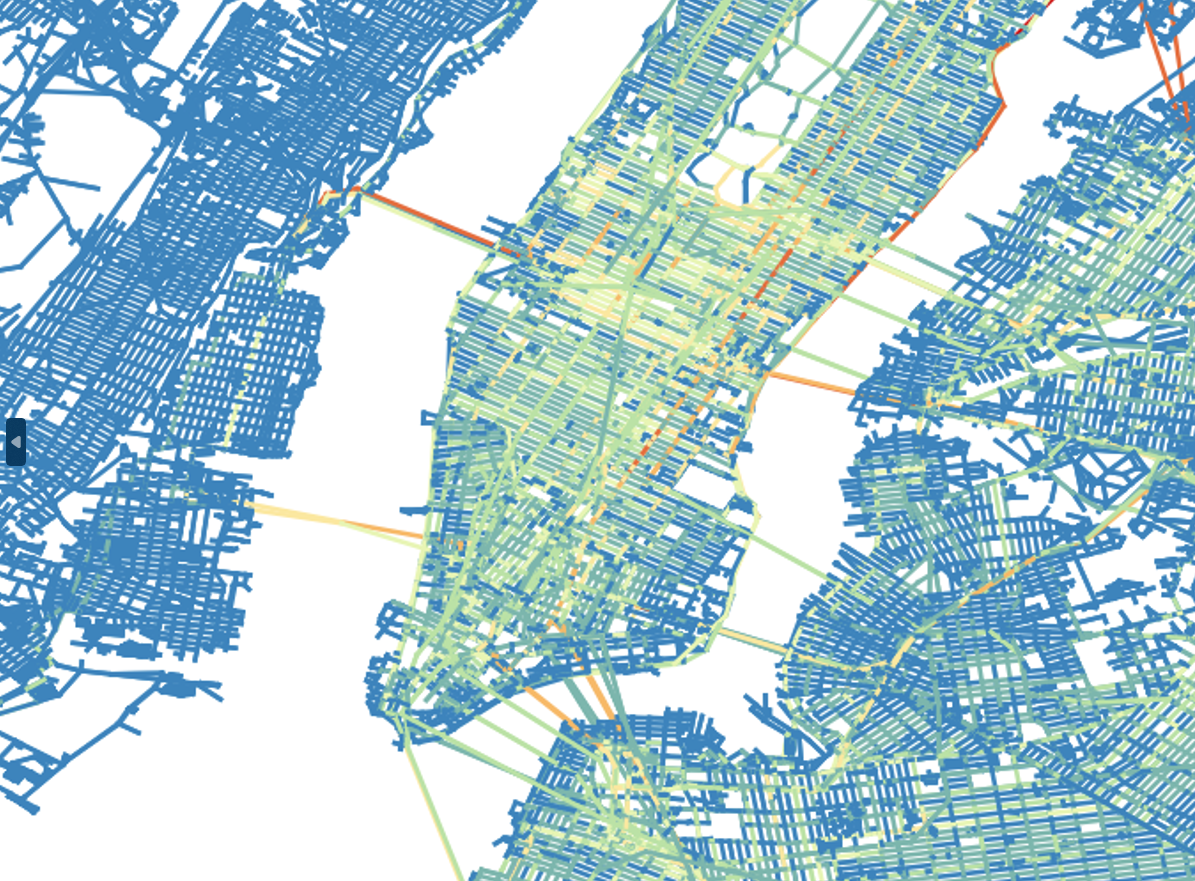}
\caption{Link congestion level during 17:00 - 18:00 with toll}
\label{withtoll}
\end{figure}

More than the link-level improvement, we also evaluate the performance of the bench-marked pricing strategy at a system level. We utilized the software QGIS \cite{QGIS_software} with the obtained link statistics to visualize the NYC network. The link-level traffic flow is divided into five specific periods: 8:00 - 9:00, 11:00 - 12:00, 12:00 - 13:00, 17:00 - 18:00, and 20:00 - 21:00. The pairwise comparison is shown in Figure \ref{withouttoll} and Figure \ref{withtoll}. It can be noticed that the traffic congestion is qualitatively improved. We can find that the congestion during rush hours has been mitigated, especially for those ``bottleneck" links such as bridges and tunnels, which indicates the effectiveness of the proposed pricing strategy. However, we notice that some links are undertaking more traffic flow than before, as shown in Figure \ref{withtoll}, but it makes sense because the traffic is rerouted to less congested links.

\section{Conclusion} \label{conclusion}
This study evaluates a complete tolling scheme via the agent-based simulation platform MATSim in a large-scale yet fine-granularity setting. The entire NYC traffic network (all five boroughs) model and 320k individual agents are used for the simulation. The results uncover microscopic behaviors under the introduced tolling regulation, which is informative to account for macroscopic trends and not available from traditional coarse simulations. The findings indicate that the tested Manhattan tolling scheme can reduce traffic congestion during peak hours, lower the usage of private cars, and encourage greater reliance on public transportation. Due to limited computational resources, only 4\% of the NYC population is simulated, and the simulations can be extended to more tolling policies to identify the most effective strategy that aligns with public interests. Our work showcases the feasibility and necessity of leveraging agent-based simulations for complicated systems, and we believe it will build foundations for developing simulation-based optimization algorithms, not limited to congestion control but many other downstream applications that can contribute to more sustainable transportation systems.




\bibliographystyle{IEEEtran}
\bibliography{references.bib}

\end{document}